\documentstyle[prl,aps,epsf]{revtex} \def\narrowtext{} \tighten \twocolumn
\input epsf.sty
\begin{document}
\draft

\title{Electronic Spectra and Their Relation to the $(\pi,\pi)$ Collective 
Mode in High-$T_{c}$ Superconductors}
\author {
        J. C. Campuzano,$^{1,2}$
        H. Ding,$^3$
        M. R. Norman,$^2$
        H.M. Fretwell,$^{1}$
        M. Randeria,$^4$
        A. Kaminski,$^1$
        J. Mesot,$^2$
        T. Takeuchi,$^{5}$
        T. Sato,$^6$ T. Yokoya,$^{6}$
        T. Takahashi,$^6$
        T. Mochiku,$^{7}$
        K. Kadowaki,$^{8}$
        P. Guptasarma,$^{2}$ D. G. Hinks,$^{2}$
        Z. Konstantinovic,$^{9}$ Z. Z. Li,$^{9}$ and H. Raffy$^{9}$
       }
\address{
         (1) Department of Physics, University of Illinois at Chicago,
             Chicago, IL 60607\\
         (2) Materials Sciences Division, Argonne National Laboratory,
             Argonne, IL 60439 \\
         (3) Department of Physics, Boston College, Chestnut Hill,
             MA  02467 \\
         (4) Tata Institute of Fundamental Research, Mumbai 400005,
             India\\
         (5) Department of Crystalline Materials Science, Nagoya
             University, Nagoya 464-01, Japan\\
         (6) Department of Physics, Tohoku University, 980-8578 Sendai,
             Japan\\
         (7) National Research Institute for Metals, Sengen, Tsukuba,
             Ibaraki 305, Japan\\
         (8) Institute of Materials Science, University of Tsukuba,
             Ibaraki 305, Japan\\
         (9) Laboratoire de Physique des Solides, Universite de
             Paris-Sud, 91405 Orsay Cedex, France\\
         }

\address{%
\begin{minipage}[t]{6.0in}
\begin{abstract}
Photoemission spectra of $Bi_{2}Sr_{2}CaCu_{2}O_{8+\delta}$ reveal that the
high energy feature near $(\pi,0)$, the ``hump", scales with the
superconducting gap and persists above $T_c$ in the pseudogap phase. As the
doping decreases,
the dispersion of the hump increasingly reflects the wavevector $(\pi,\pi)$
characteristic of the undoped insulator, despite the presence of a large
Fermi surface. This can be understood from the interaction of the
electrons with a collective mode, supported by our
observation that the doping dependence of the resonance observed by neutron
scattering is the same as that inferred from our data.
\typeout{polish abstract}
\end{abstract}
\end{minipage}}

\maketitle
\narrowtext

In the high temperature copper oxide superconductors,
a small change in doping takes the material from an antiferromagnetic
insulator to a d-wave superconductor.  This raises the fundamental
question of the relation of the electronic structure of the doped
superconductor \cite{DING96}
to that of the parent insulator \cite{WELLS}.
Here we examine this by using angle resolved photoemission spectroscopy
(ARPES). We find that the spectral lineshape and its dispersion
evolves as a function of doping
from one which resembles a strong coupling effect of superconductivity
in the overdoped limit to one which resembles the
insulator in the underdoped limit.  The connection between
these two limits can be understood in terms of a collective mode
\cite{ROSSAT,MOOK} which has the same $(\pi,\pi)$ wavevector
characteristic of the magnetic insulator, and whose energy decreases
as the doping is reduced.  This is supported by our
observation that the mode energy inferred from ARPES
as a function of doping correlates strongly with that obtained directly
from neutron scattering data \cite{BOURGES}, and points to the intimate
relation of magnetic correlations to high $T_{c}$ superconductivity.

The experiments were carried 
out using procedures and samples described previously \cite{NATURE1}, 
as well as films grown by RF magnetron sputtering \cite{FILM}.
The doping level was controlled by varying oxygen stochiometry, with
samples labeled by their onset $T_{c}$. Spectra were obtained
with a photon energy of 22 eV and a photon polarization
directed along the CuO bond direction. Spectra had energy resolutions 
(FWHM) of 17, 26, or 34 meV with a momentum window of radius 0.045$\pi$/a. 
Energies are measured with respect to the chemical potential, determined 
using a polycrystalline Pt or Au reference in electrical contact with the
sample.

We begin with the $T$ evolution of the spectra of
$Bi_{2}Sr_{2}CaCu_{2}O_{8+\delta}$ (Bi2212) near the
$(\pi,0)$ point of the Brillouin zone (inset of Fig.~1a). In
the underdoped region of the phase diagram (that lies
between the undoped insulator and optimal doping, corresponding
to the highest $T_c$), one observes a
pseudogap ($30-50$ meV) which is very likely associated
with pairing above $T_c$, a precursor to superconductivity
\cite{MARSHALL,NATURE1}. For temperatures above the pseudogap
temperature scale $T^*$ \cite{NATURE1} we see a broad peak
which is chopped off by the Fermi function, as shown in
Fig.~1a for an underdoped 83K sample at 200 K.
In this respect, the one-particle spectral function of
the underdoped compounds, which is completely incoherent,
is similar to that observed in the overdoped compounds. While
there is only weak dispersion from $(\pi,0) \to (\pi,\pi)$
for $T > T^*$, there is definite loss of integrated
spectral weight \cite{NK} and one can identify the
$(\pi,0) \to (\pi,\pi)$ ``Fermi surface'' crossing \cite{DING97}.

As the temperature is reduced below $T^{*}$, but still above $T_c$, we see
that the spectral function remains completely incoherent, as
shown in Fig.~1b for an underdoped 89K sample.  The
leading edge pseudogap\cite{MARSHALL,NATURE1} which develops below
$T^{*}$ is difficult to see on the energy scale of Fig.~1b (the 
midpoint shift at 135K is 3 meV). However, a 
higher energy feature (the ``high energy pseudogap'') can easily be 
identified by a change in slope of the spectra as a function of 
energy (see Fig.~2). On further reduction of the temperature below 
$T_c$, a coherent quasiparticle
peak begins to grow at the position of the leading edge gap,
accompanied by a redistribution of the incoherent spectral
weight leading to a dip and hump structure \cite{DESSAU,NK}. The 
peak-dip-hump lineshape and the dispersion of these features will play 
a central role in our discussion. 

The high energy pseudogap feature is closely related to the 
hump below $T_c$, as seen from a comparison of their dispersions. We show 
data along $(\pi,0) \to (\pi,\pi)$ for an underdoped 75K sample in the 
superconducting state (Fig.~2a) and in the pseudogap regime (Fig.~2b). 
Below $T_c$, the sharp peak at low energy is essentially dispersionless, 
while the higher energy hump rapidly disperses from
the $(\pi,0)$ point towards the $(\pi,0) \to (\pi,\pi)$ Fermi crossing
\cite{DING97} seen above $T^*$.  Beyond this, the intensity drops dramatically,
but there is clear evidence that the hump disperses back to higher 
energy.  In the pseudogap state, the high energy feature also shows 
strong dispersion \cite{MARSHALL,WHITE}, much like the
hump below $T_{c}$, even though the leading edge is non-dispersive
like the sharp peak in the superconducting state.

In Fig.~3 we show the dispersion of the sharp peak and hump
(below $T_c$), for a variety of doping levels,
in the vicinity of the $(\pi,0)$ point along the two principal axes.
The sharp peak at low energies is seen to be essentially
non-dispersive along both directions for all doping levels, while
the hump shows very interesting dispersion.
Along $(\pi,0) \to (0,0)$ (Fig.~3a), the hump exhibits
a maximum, with an eventual dispersion away from the Fermi energy,
becoming rapidly equivalent to the binding energy of the broad peak in
the normal state as one moves away from the region near $(\pi,0)$\cite{MODE1}.
In the orthogonal direction (Fig.~3b), since the hump
initially disperses towards the $(\pi,0) \to (\pi,\pi)$
Fermi crossing, which is known to be a weak function of
doping\cite{DING97}, one obtains the rather dramatic effect that the
dispersion becomes stronger with underdoping.
We also note that there is an energy separation between
the peak and the hump due to the spectral dip. In essence, the hump
disperses towards the spectral dip, but cannot cross it, with its
weight dropping strongly as the dip energy is approached.  Beyond this point,
one sees evidence of the dispersion bending back to higher binding energy for
more underdoped samples.

Fig.~4a shows the evolution of the low temperature spectra at the $(\pi,0)$ 
point as a function of doping. The sharp quasiparticle peak moves to higher 
energy, indicating that the gap increases with underdoping \cite{HARRIS} 
(although this is difficult to see on the scale of Fig.~4a). We see that 
the hump moves rapidly to higher energy with underdoping \cite{LAUGHLIN}. 
These trends can be seen very clearly in Fig.~4b, where the energy of the 
peak and hump are shown as a function of doping for a large number of samples.
Finally, we observe that the quasiparticle peak loses spectral weight with 
increasing underdoping, as expected for a doped Mott insulator; in addition 
the hump also loses spectral weight though less rapidly.

The hump below $T_c$ is clearly related to the
superconducting gap, given the weak doping
dependence of the ratio between the hump and quasiparticle peak positions at
$(\pi,0)$, shown in Fig.~4c.
Tunneling data find this same correlation on a wide variety of
high-$T_{c}$ materials whose energy gaps vary by a factor of 
30 \cite{JOHNZ}. We have additional strong evidence \cite{DING96,MODE1} 
that the peak and hump do {\it not} arise from two different ``bands''.

Thus, the peak, dip and hump are features of a single spectral 
function, and imply a strong frequency dependence of the superconducting 
state self-energy (a ``strong-coupling effect'').
The hump represents the energy scale at which  
the spectral function below $T_c$ matches onto that in the 
normal state (as evident from the data in the bottom curve of Fig.~1b). 
However, the existence of the dip requires additional structure in 
the self-energy. We had suggested that this structure can 
be naturally understood in terms of electrons interacting with
a sharp collective mode \cite{MODE1,MODE2} below $T_c$, which also 
leads to an explanation of the non-trivial dispersion, as 
discussed below. It was speculated that the mode was
the same as that observed directly by neutron scattering in
$YBa_2Cu_3O_7$ \cite{ROSSAT,MOOK},
and more recently in Bi2212 \cite{MOOK2,KEIMER2}.

To motivate the analysis below that firmly establishes
the mode interpretation of ARPES spectra and its connection
with neutron data, we need to recall \cite{MODE1,MODE2}
that the spectral dip represents a pairing induced gap in the incoherent
part of the spectral function at $(\pi,0)$ occurring at an energy
$\Delta + \Omega_0$,  where $\Delta$ is the superconducting gap and
$\Omega_0$ is the mode energy. We can estimate the mode energy from 
ARPES data from the energy difference between the dip ($\Delta + \Omega_0$)
and the quasiparticle peak ($\Delta$).

In Fig.~5b we plot the mode energy as estimated from ARPES for various
doping levels as a function of $T_c$ and compare it with neutron measurements.
We find striking agreement both in terms of the energy scale
and its doping dependence \cite{BOURGES}. We note that the mode energy inferred
from ARPES decreases with doping, just like the
neutron data, unlike the gap energy (Fig.~4b), which increases.
This can be seen directly in the raw data, shown in Fig.~5a.
Moreover, there is strong correlation between the temperature
dependences in the ARPES and neutron data.  While neutrons see a sharp mode
only below $T_c$, a smeared out remnant persists up to $T^*$ \cite{MOOK3}.
As the sharpness of the mode is responsible for the sharp spectral dip, one
then sees the
correlation with ARPES where the dip disappears above $T_c$, but with
a remnant of the hump persisting to $T^*$.

An important feature of the neutron data is that the mode only exists
in a narrow momentum range about $(\pi,\pi)$, and is magnetic in
origin\cite{MOOK}.  To see a further connection
with ARPES, we return to the results of Fig.~3.  Note the dispersion
along the two orthogonal directions are similar (Fig.~3c),
unlike the dispersion inferred in the normal state\cite{DING96}.
As these two directions are related by a $(\pi,\pi)$ translation
($(x,0)\equiv (0,-x); (0,-x)+(\pi,\pi) = (\pi,\pi-x)$), we see that the
hump dispersion is clearly reflecting the $(\pi,\pi)$ nature of the
collective mode.  This dispersion is also consistent
with a number of models \cite{KAMPF,SDW1}
in the literature which identify the high energy
feature in the pseudogap regime as a remnant of the insulating magnet.
We note, though, that the mode is due to quasiparticle pair creation and thus
not just a continuation of the spin wave mode from the
antiferromagnet \cite{DEMLER1}.

This brings up a question that is at the heart of the high $T_c$ problem:
how can a feature which can be understood as a strong coupling effect
of superconductivity, as discussed above, turn out to have a dispersion that
resembles
that of a magnetic insulator?  The reason is that the collective mode
has the same wavevector, $(\pi,\pi)$, which characterizes the magnetic order
of the insulator.
It is easy to demonstrate that in the limit that the mode energy goes
to zero (long range order), one actually reproduces a
symmetric dispersion similar to that in Fig.~3c, with the spectral gap 
determined by the strength of the mode\cite{KAMPF}.  This is in 
accord with the
increase in the hump energy with underdoping (Fig.~4b) tracking the 
rise in the neutron mode intensity\cite{BOURGES}.
Since the hump scales with the superconducting gap,
the obvious implication is that the mode is intimately connected with
pairing, a conclusion which can also be made by relating the mode to the
superconducting condensation energy \cite{DEMLER2}.  That is, high $T_c$
superconductivity is likely due to the same magnetic correlations which
characterize the insulator and give rise to the mode.

This work was supported by the National Science Foundation DMR 9624048, and
DMR 91-20000 through the Science and Technology Center for Superconductivity,
the U. S. Dept. of Energy, Basic Energy Sciences, under contract
W-31-109-ENG-38, the CREST of JST, and the Ministry of
Education, Science, and Culture of Japan. The Synchrotron Radiation Center is
supported by NSF DMR 9212658. JM is supported by the Swiss National Science
Foundation, and MR by the Swarnajayanti fellowship of the Indian DST.

\begin{figure}
\epsfxsize=3.4in
\epsfbox{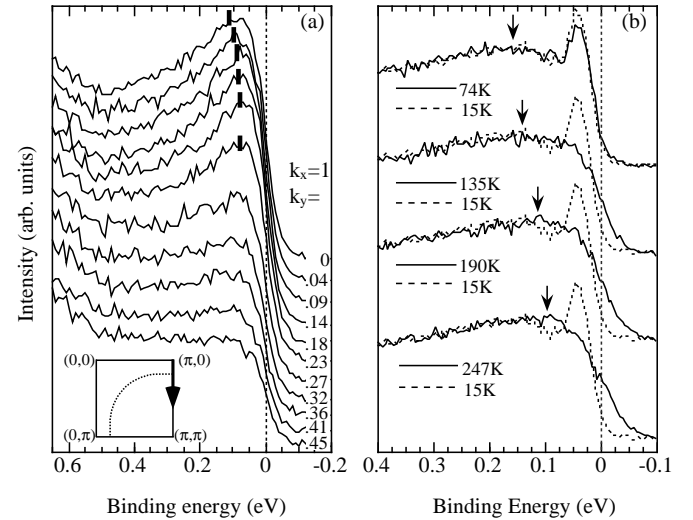}
\vspace{0.5cm}
\caption{Dispersion and temperature dependence of spectra near 
$(\pi,0)$:
(a) Spectra along $(\pi,0) \to (\pi,\pi)$
for an underdoped 83K sample at 200K (above $T^*$), with the thick
vertical bar indicating the peak position.  The curves are labeled in
units of $\pi/a$.  The Brillouin zone is shown as an inset, with the Fermi
surface as a dotted line.  (b) Temperature evolution of the spectra at the
$(\pi,0)$ point for an underdoped 89K sample, with the positions of 
the high energy feature marked by arrows.}
\label{fig1}
\end{figure}

\begin{figure}
\epsfxsize=3.4in
\epsfbox{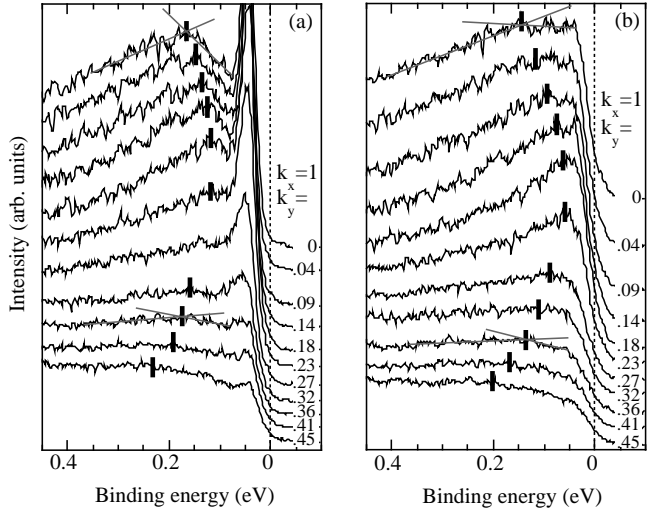}
\vspace{0.5cm}
\caption{
Spectra along $(\pi,0) \to (\pi,\pi)$ in (a)
the superconducting state (T=60K), and (b) the pseudogap state
(T=100K) for an underdoped 75K sample (curves are labeled in units of
$\pi/a$). The thick vertical bar indicates the position of the higher
energy feature, at which the spectrum changes slope as highlighted by 
the intersecting straight lines.}
\label{fig2}
\end{figure}

\begin{figure}
\epsfxsize=3.4in
\epsfbox{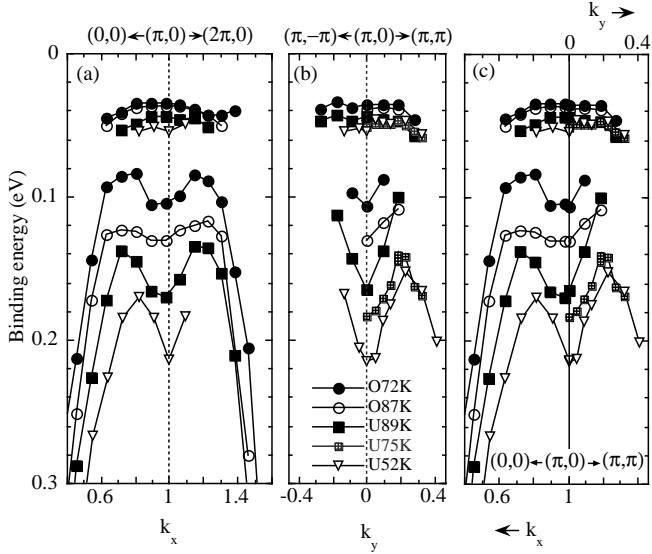}
\vspace{0.5cm}
\caption{
Doping dependence of the dispersion from (a) $(\pi,0) \to (\pi\pm\pi,0)$, 
(b) $(\pi,0) \to (\pi,\pm\pi)$, and (c) both directions, for the
peak and hump in the superconducting state. U is underdoped and O is
overdoped. Points were obtained by polynomial fits to the data, and 
are consistent with the simpler criterion used in Fig.~2.}
\label{fig3}
\end{figure}

\begin{figure}
\epsfxsize=3.4in
\epsfbox{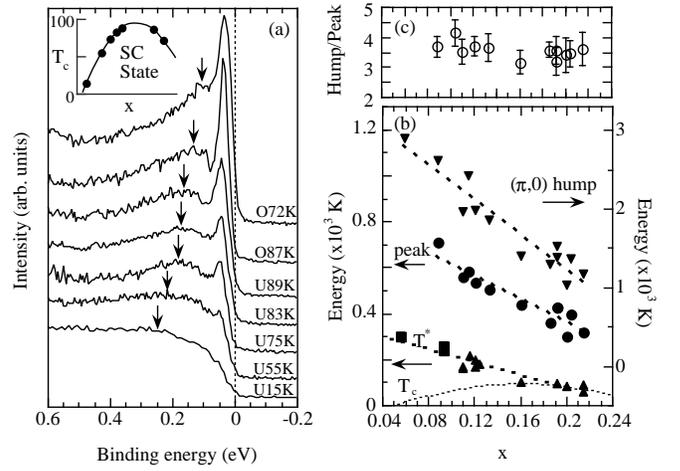}
\vspace{0.5cm}
\caption{Dependence of energy scale on carrier density:
(a) Doping dependence of the spectra (T=15K) at
the $(\pi,0)$ point (U55K is a film). The inset shows $T_c$ vs. doping.
(b) Doping dependence of $T^{*}$, and the peak and hump binding energies
in the superconducting state along with their
ratio (c), as a function of doping, $x$. The empirical relation between $T_c$
and $x$ is given by $T_c/T_c^{max}=1-82.6(x-0.16)^2$ \protect\cite{PRES} with
$T_c^{max}$=95K. For $T^{*}$, solid squares represent lower bounds.}
\label{fig4}
\end{figure}

\begin{figure}
\epsfxsize=3.4in
\epsfbox{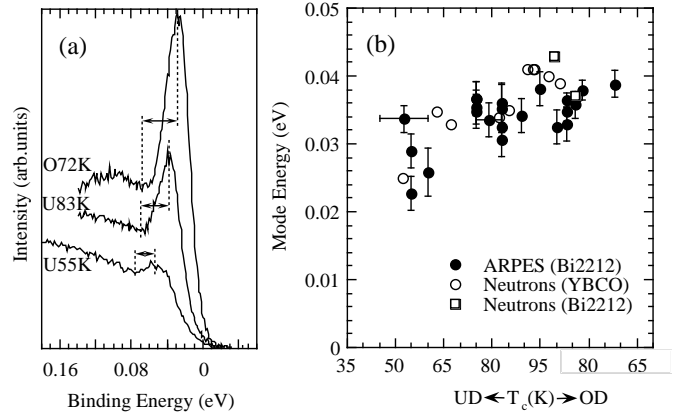}
\vspace{0.5cm}
\caption{Doping dependence of the mode energy:
(a) Spectra at $(\pi,0)$ showing the decrease in the energy separation
of the peak and dip with underdoping. Peak and dip locations were 
obtained by independent polynomial fits and carefully checked for the 
effects of energy resolution.
(b) Doping dependence of the collective mode energy inferred from ARPES
together with that inferred from neutron data (for latter, YBCO results as
compiled in Ref.~\protect\onlinecite{BOURGES}, Bi2212 results of
Refs.~\protect\onlinecite{MOOK2} and \protect\onlinecite{KEIMER2}).}
\label{fig5}
\end{figure}

\end{document}